\documentclass[preprint2]{aastex}

\shorttitle{Metal-mass-to-light ratio of the Perseus cluster}
\shortauthors{Matsushita et al.}

\begin{document}


\title{Metal-mass-to-light ratios of the Perseus cluster out to the virial radius}


\author{K. Matsushita\altaffilmark{1}, E. Sakuma\altaffilmark{1}, 
T. Sasaki\altaffilmark{1},  K. Sato\altaffilmark{1} and A. Simionescu\altaffilmark{2}}

\altaffiltext{1}{Department of physics, Tokyo University of Science, 1-3 Kagurazaka, Shinjuku-ku, Tokyo 162-8601, Japan; matusita@rs.kagu.tus.ac.jp}
\altaffiltext{2}{KIPAC, Stanford University, 452 Lomita Mall, Stanford, CA 94305, USA}
\altaffiltext{3}{Department of Physics, Stanford University, 382 Via Pueblo Mall, Stanford, CA 94305-4060, USA}

\begin{abstract}
We analyzed XMM-Newton data of the Perseus cluster out to $\sim$1 Mpc, 
or approximately half the virial radius. Using the flux ratios of Ly$\alpha$ 
lines of H-like Si and S   to K$\alpha$ line of  He-like Fe,  the
abundance ratios of Si/Fe and S/Fe of the intracluster medium
(ICM) were derived using the  APEC plasma code v2.0.1.
The temperature dependence of the line ratio limits the systematic uncertainty 
in the derived abundance ratio.
The Si/Fe and S/Fe in the ICM of the Perseus cluster show no radial gradient.
The emission-weighted averages of the Si/Fe and S/Fe ratios outside the cool core
 are 0.91 $\pm$ 0.08 and 0.93 $\pm$ 0.10, respectively,
in solar units according to the solar abundance table of \citet{lod03}.
These ratios indicate that  most  Fe was synthesized by supernovae Ia.
We collected $K$-band luminosities of galaxies and calculated the ratio 
of Fe and Si mass in the ICM
 to $K$-band luminosity, iron-mass-to-light ratio (IMLR) and 
silicon-mass-to-light ratio (SMLR).
Within $\sim$1 Mpc, the cumulative IMLR and SMLR increase with radius.
Using Suzaku data for the  northwest and east directions,
we also calculated the IMLR out to $\sim$ 1.8 Mpc, or about the virial radius.
We   constrained the SMLR out to this radius and discussed the slope of
the  initial mass function of stars in the cluster.
Using the cumulative IMLR profile, we discuss the past supernova Ia rate.

\end{abstract}


\section{Introduction}

Clusters of galaxies are the largest gravitationally bound objects in the Universe.
The intracluster medium (ICM) contains a large amount of metals,
synthesized by supernovae (SNe) in cluster galaxies.
Thus, the distribution of metals
 in the ICM  provides important information
on the chemical history and evolution of clusters.

Because metals were synthesized in
galaxies, the ratios of metal mass in the ICM to the total light
from galaxies in clusters or groups, (i.e., metal-mass-to-light ratios)
are key parameters in investigating the chemical evolution of the
ICM\@.
Using Einstein and Ginga data, \citet{Arnaud1992} and \citet{Tsuru1992}
derived ratios of Fe mass in the ICM to the total light from galaxies, which 
is the  iron-mass-to-light ratio (IMLR).
To account for the observed IMLR,  either the past average rate of SN Ia
 was at least higher by a factor of 10 than the present rate or massive stars in 
clusters formed with a very flat initial mass function \citep{Renzini1993}.
With ASCA observations,
the derived  IMLR  within a radius where the ICM density
falls below 3 $\times 10^{-4} ~\rm{cm^{-3}}$
 is nearly constant in rich clusters and decreases toward poorer systems \citep{Makishima2001}. 
With  Suzaku, Chandra, and  XMM satellites, 
the IMLR of several medium-sized clusters 
and groups of galaxies was measured out to 0.2$r_{180}$--0.5$r_{180}$ 
\citep{matsushita07a, komiyama09, Sato2007b, sato08, 
sato09a, sato09b, sato10,  Rasmussen2009,  sakuma11} and that of the 
Coma cluster  was derived out to 0.5$r_{180}$ \citep{Matsushita2012}.
Suzaku first measured the Fe abundance of the ICM 
beyond 0.5$r_{180}$ \citep{fujita08, Simionescu2011}.
With Suzaku, \citet{sato12} derived the IMLR profile of the Hydra A
cluster out to $r_{180}$.
Here, $r_{180}$ is the radius
  in which matter at 180 times the critical density of the universe is contained.
In individual clusters, the IMLR is lower around the center,
indicating that Fe in the ICM extends farther than stars. 
Therefore, to derive the total Fe mass in the ICM, we need observations out
to the virial radius.

Since Fe is synthesized by both SNe Ia and by core-collapse SNe (hereafter SNecc), 
we need measurements of abundances of various elements 
to constrain contributions from the two types of SNe.
The ASCA satellite first studied the Si abundance in the ICM \citep{Fukazawa1998, fukazawa00, Finoguenov2000, Finoguenov2001}.
\citet{Fukazawa1998} 
 reported that the Si/Fe ratio in the ICM increases with ICM temperature, and
\citet{Finoguenov2000} reported that the Si/Fe  ratio increases
with radius in several clusters.    
Using Chandra data of groups out to $r_{500}$, 
\citet{Rasmussen2007} reported that the SNcc contribution increases with radius and completely dominates at  $r_{500}$.
 XMM-Newton and Suzaku observations have also been used to study
the Si/Fe ratio of the ICM in clusters and  groups of galaxies.
\citep[e.g.][]{Matsushita2003, Tamura2004,  dePlaa2007, Rasmussen2007, matsushita07a, Matsushita2007b,  Sato2007b, sato08, komiyama09,  simonescu09, deGrandi2009,  sato09a, sato09b, sato10, sakuma11}.
 With Suzaku observations of clusters and groups with  ICM temperatures lower than
$\sim4 $ keV,   the derived values of Si abundance 
 are close to those of Fe out to 0.2$r_{180}$--0.3$r_{180}$, 
with a small scatter when using the solar abundance table by \citet{lod03}.
 On the basis of the  abundance ratios of  Si and Fe,
 the contributions from SN Ia and SNcc have been estimated.
However, excluding cool-core regions,
the error bars in the Si/Fe ratio of hotter clusters 
are very large \citep{Tamura2004}.
\citet{Matsushita2012}
derived the Si/Fe ratio in the ICM of the Coma cluster from the 
 flux ratios of 
the Ly$\alpha$ line of H-like Si and the K$\alpha$ line of He-like Fe.
 The temperature dependence of the line ratio above several keV is relatively small
and  limits the systematic uncertainty in the derived abundance ratio.
The derived Si/Fe ratio of the Coma cluster is approximately 1 solar
according to the same solar abundance table,  with no radial gradient out to 0.5$r_{180}$.

The Perseus cluster  is the brightest cool-core cluster with a redshift of 0.018.
With Chandra observations, the cool core of the Perseus cluster shows
complicated features such as ripples and
 shocks around radio bubbles \citep{Fabian2006}.
With XMM, \citet{Tamura2004} derived the Si/Fe ratio outside the cool core of the Perseus cluster, (50--200 $h_{100}^{-1}$ kpc), to be
 0.77 $\pm$ 0.25  in solar units using the solar abundance table by \citet{lod03}.
Here, the Hubble constant is $H_0= 100 h_{100}$~km~s$^{-1}$~Mpc$^{-1}$.
With Suzaku data, 
\citet{tamura2009} derived the abundance distribution of Mg, Si, S, Ar, Ca, Cr, Mn,  Fe,
and Ni of the central region of the Perseus cluster.
\citet{Simionescu2011} derived the electron density and Fe abundance profiles out to 
the virial radius toward the northwest (NW) and east (E) directions
observed with Suzaku.
\citet{Simionescu2012} studied the large-scale motions of the ICM of the 
Perseus cluster using ROSAT, XMM,  and Suzaku data.

In this paper, we study the Si/Fe and S/Fe ratios in the ICM of the Perseus cluster 
observed with XMM out to 1.1~Mpc.
We collect the $K$-band luminosity of galaxies and calculated the IMLR and
Si-mass-to-light ratios (SMLR) out to this radius.
We also calculated the  IMLR and constrained SMLR
 out to  1.8~Mpc, or about the virial radius,
using the Suzaku measurements in the  NW and E directions by \citet{Simionescu2011}.
Using a Hubble constant of  70~km~s$^{-1}$~Mpc$^{-1}$, 
 1$'$ corresponds to 22 kpc\@.
The value of $r_{180}$ expected from numerical simulations is 
 $r_{180}=1.95~h_{100}^{-1}\sqrt{k \langle T \rangle/10~\rm{keV}}~\rm{Mpc}$
\citep{Evrard1996, Markevitch1998}.
For the Perseus cluster,  this value  corresponds to  2.2~Mpc 
using the average temperature $k \langle T \rangle$=6.1 keV.
Hereafter, we denote this radius  as $r_{180\rm<kT>}$.
An fit with the universal mass model of 
Navarro, Frenk and White (NFW)  \citep{navarro1996, navarro1997} 
 to the hydrostatic mass derived from the Suzaku observations
in the NW direction gave $r_{200}$ (hereafter $r_{200\rm<HE>}$)  as 1.8 Mpc
\citep{Simionescu2011}.

We used the solar abundance table by \citet{lod03},
in which the solar Si, S and  Fe abundances relative to H are
3.47$\times$10$^{-5}$,  1.55$\times$10$^{-5}$, and 2.95$\times$10$^{-5}$,
respectively,  by number.
Considering a difference in solar He abundance, 
the Fe abundance yielded by \citet{lod03}  
is 1.5 times higher than that using the photospheric value by \citet{angr}.
Using the table by \citet{lod03}, the Si/Fe and S/Fe ratios 
are factors of 1.55 and 1.52, respectively,  smaller than those from \citet{angr}.

In this paper,
errors were quoted at a 68\% confidence level for a single parameter.

\section{Observations}

\begin{deluxetable}{ccccc}
  \tablecaption{Observation log of the Perseus cluster}
\tablewidth{0pt}
\tablehead{
\colhead{obsid\tablenotemark{a}} &\colhead{ (RA, Dec)\tablenotemark{b}}&\colhead{ Date\tablenotemark{c}} &\colhead{  Exposures (ks) \tablenotemark{d}} 
} 
\startdata
0085110101&03$^{\rm h}$  19$^{\rm m}$ 48.36$^{\rm s}$  +41$^{\circ}$ 30$'$ 40.6$''$ & 2001-01-30 & 49, 51, 48\\
0085590201&03$^{\rm h}$ 19$^{\rm m}$ 45.12$^{\rm s}$ +41$^{\circ}$ 05$'$ 02.5$''$  & 2001-02-10& 42, 40, 38 \\
0151560101& 03$^{\rm h}$ 16$^{\rm m}$ 39.07$^{\rm s}$	+41$^{\circ}$  18$'$ 38.5$''$  & 2003-02-26 & 25, 25, 18 \\
0204720101&03$^{\rm h}$ 21$^{\rm m}$ 33.80$^{\rm s}$	+41$^{\circ}$  31$'$ 03.1$''$  &	2004-02-04 & 15, 15, 12 \\
0204720201	&	 03$^{\rm h}$ 23$^{\rm m}$ 18.92$^{\rm s}$	+41$^{\circ}$  30$'$ 57.2$''$  &	2004-02-04 & 25, 25, 21\\
0305720301	&	03$^{\rm h}$ 22$^{\rm m}$ 16.00$^{\rm s}$	+41$^{\circ}$  11$'$ 28.0$''$  &	2005-08-03 & 20, 21, 16\\
0305720101	&	03$^{\rm h}$ 18$^{\rm m}$ 01.94$^{\rm s}$	+41$^{\circ}$  46$'$ 46.2$''$  &	2005-09-01 & 13, 13, 10\\
0305690101	&	03$^{\rm h}$ 17$^{\rm m}$ 58.31$^{\rm s}$	+41$^{\circ}$  16$'$ 15.7$''$  &	2006-02-10 & 26, 27, 26 \\
0305690301	&	03$^{\rm h}$ 19$^{\rm m}$ 45.91$^{\rm s}$	+41$^{\circ}$  52$'$ 51.0$''$  &	2006-02-11  & 20, 20, 16\\
0305690401 &	03$^{\rm h}$ 21$^{\rm m}$ 49.07$^{\rm s}$	+41$^{\circ}$  48$'$ 47.2$''$  &	2006-02-11 & 27, 27, 25 \\

0405410101	&	03$^{\rm h}$ 21$^{\rm m}$ 04.82$^{\rm s}$	+41$^{\circ}$  56$'$ 46.7$''$  &	2006-08-03 & 18, 16, 12\\
0405410201	&	03$^{\rm h}$ 18$^{\rm m}$ 44.75$^{\rm s}$	+41$^{\circ}$  07$'$ 11.2$''$ &	2006-08-03 & 17, 26, 9\\
\enddata
\label{tab:obslog}
\tablenotemark{a}{XMM-Newton observational identifier }
\tablenotemark{b}{in J2000.0}
\tablenotemark{c}{Date of start of the observation}
\tablenotemark{d}{Exposure times of MOS1, MOS2, and PN, respectively,
after screenings}
\end{deluxetable}
\begin{figure}[t]
\includegraphics[width=8cm, clip=]{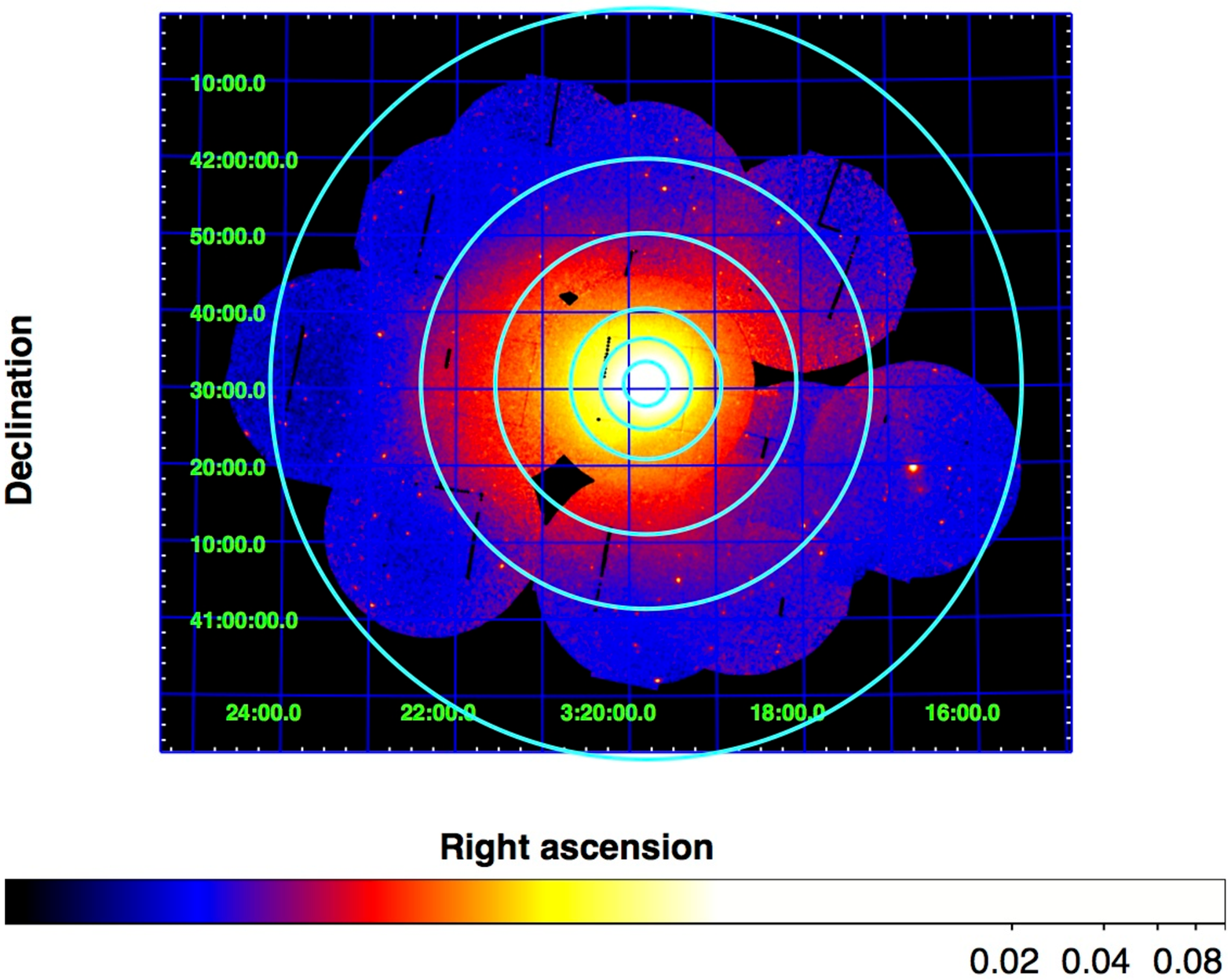}
\caption{Exposure-corrected combined MOS  image of the Perseus cluster
(0.5--4.0 keV). 
The circles have radii of 0.03, 0.06, 0.1,  0.2, 0.3, and 0.5$r_{180\rm<kT>}$. 
The numbers below the color bar have units of  counts $\rm{s^{-1} pixel^{-1}}$.
}
\label{fig:xmmimage}
\end{figure}

We analyzed the archival data of the XMM-Newton observations 
of the Perseus cluster 
using PN, MOS1, and MOS2 detectors.
In this study, we used SASv12.0, but the
 details of observations, event selection, and  background subtraction
  are the same as those in \citet{Matsushita2011}.
The observation log is summarized in Table \ref{tab:obslog}.
We included three observations in the archive  in addition to
those in \citet{Matsushita2011}.
The exposure-corrected combined MOS  image of the Perseus cluster
within an energy range of 0.5--4.0 keV is shown in Figure \ref{fig:xmmimage}.
Spectra were accumulated in concentric annular regions
centered on the X-ray peak of the Perseus cluster.
The spectra from MOS1 and MOS2 were added.







\section{Data Analysis and Results}
\label{sec:analysis}


\subsection{Abundance ratios of Si/Fe and S/Fe out to 0.5$r_{180\rm<kT>}$}


\begin{figure*}
\epsscale{2.0}
\includegraphics[width=16cm, clip=]{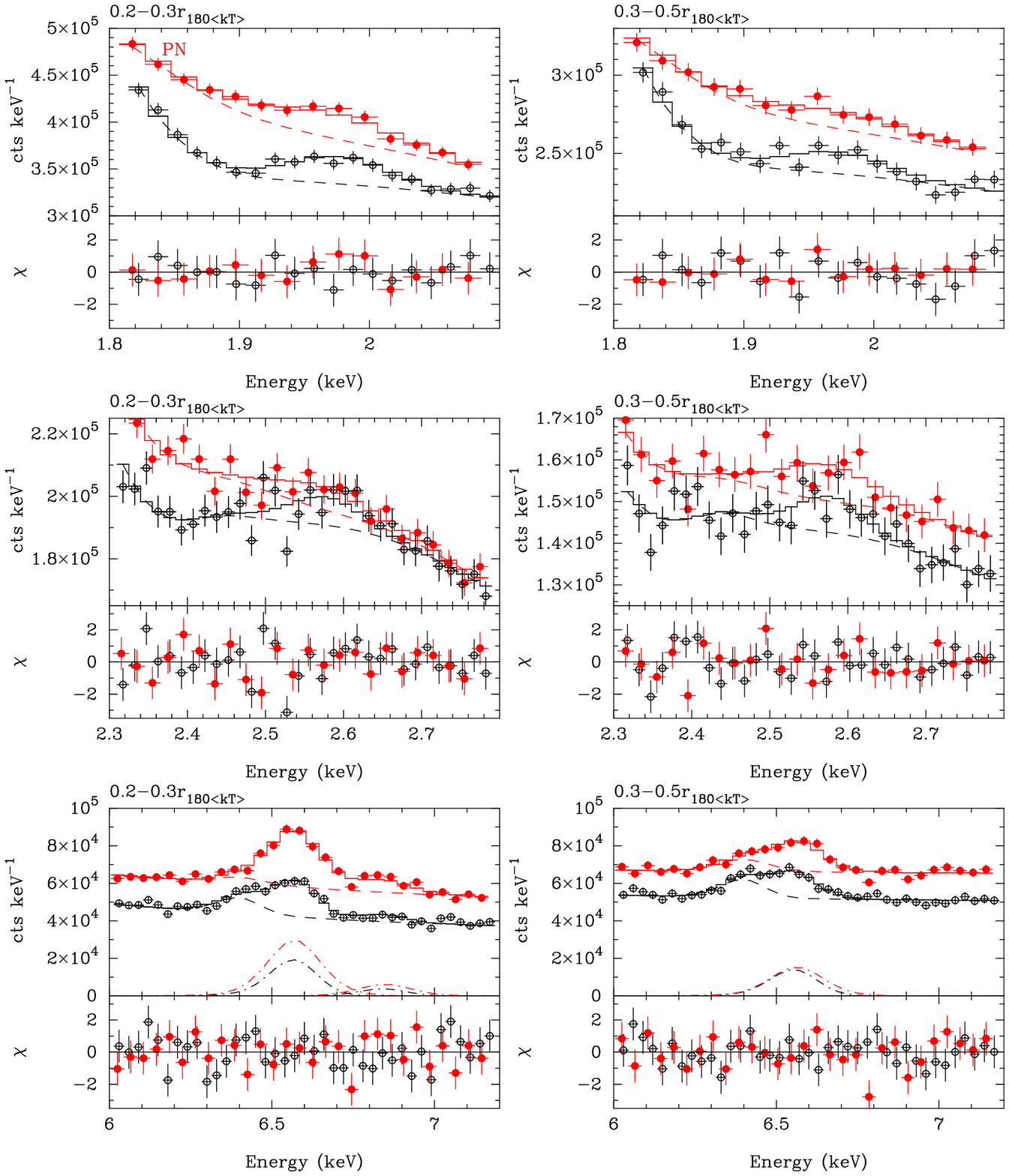}
\caption{ Spectra  at 440--650 kpc or 0.2--0.3$r_{180\rm<kT>}$ (left panels)
and 650--1090 kpc or 0.3--0.5$r_{180<\rm kT>}$ (right panels) around Ly$\alpha$ lines of H-like Si (top panels)
and S (middle panels) and K$\alpha$ line of He-like Fe (bottom panels).
Open (black) and closed (red) circles correspond to the MOS and PN spectra, respectively.
Here, the background was not subtracted.
The solid lines correspond to the best-fit model, which is a sum of 
a vAPEC v2.0.1 model, Gaussians,  and  background components.
The dashed lines show the contribution of the continuum of the ICM component
and the background.
The dot-dashed lines in the bottom panels show the contribution of the Fe lines from the ICM.
The bump at 6.4 keV in the background component corresponds to
 an instrumental line.
} 
\label{fig:sispec}
\end{figure*}

Figure \ref{fig:sispec} shows representative MOS and PN spectra around the Ly$\alpha$ lines
of H-like Si and S and K$\alpha$ line of He-like Fe.
Although the Si and S lines are clearly seen in the spectra, the peak level of the lines
in the spectra is only several percentage points above that of the continuum.
For the spectra beyond several hundred  kpc, around the Fe line, most of the continuum
comes from  non-X-ray background (NXB).
As a result, a small systematic uncertainty in the response matrix and background
can cause a large systematic uncertainty in the abundance of these elements.
For example, when we fitted MOS and PN spectra at 0.1--0.2$r_{180<\rm kT>}$
 using an energy range of 0.5--10.0 keV
with a vAPEC model  \citep{smith01}, there are discrepancies of a few percent around the Si and S lines  between the data and the model and
the derived abundances from the PN and MOS detectors differ by a factor of two.
When we restricted energy ranges around H-like lines of Si and S and refitted
the spectra, the  derived abundances strongly depend on the adopted energy range.

\begin{deluxetable}{lccccl} 
\tablecaption{Results of the spectral fitting around K$\alpha$ line of Fe}
\tablewidth{0pt}
\tablehead{
\colhead{radius} & \colhead{$F_{\rm 6.9}/F_{\rm 6.7}$ \tablenotemark{a}}& \colhead{$kT $ \tablenotemark{b}} &\colhead{$\chi^2/d.o.f$\tablenotemark{c}} &  \colhead{Fe  \tablenotemark{d}} 
\\
\colhead{(kpc/$r_{180<kT>}$)} & & (keV) & & \colhead{(solar)}}
\startdata
60--130/0.03--0.06 &  0.23 $\pm$ 0.01 & 5.9 $\pm$ 0.1 & 76/93   & 0.63 $\pm$ 0.01\\
130--220/0.06--0.1 & 0.30 $\pm$ 0.02 & 6.5 $\pm$ 0.2& 77/93     & 0.54 $\pm$ 0.03\\
220--440/0.1--0.2 & 0.27 $\pm$ 0.03 & 6.2 $\pm$ 0.2  & 128/124 & 0.47 $\pm$ 0.03\\
440--650/0.2--0.3 & 0.23 $\pm$ 0.05 & 5.9 $\pm$ 0.5   & 59/52 & 0.38 $\pm$ 0.04\\
650--1090/0.3--0.5 & ---           & 5.6 $\pm$ 0.6\tablenotemark{e} &  45/52 & 0.48 $\pm$ 0.07\\
\enddata
\label{tab:fegauss}
 \tablenotemark{a}{
Ratio of flux in units of photons cm$^{-2} \rm{s}^{-1}$ of Ly$\alpha$ line of H-like Fe and K$\alpha$ line of He-like Fe.}
 \tablenotemark{b}{The ICM temperature derived from the Fe line ratios using the theoretical expectation by APEC v2.0.1.}
\tablenotemark{c}{ $\chi^2$ and degrees of freedom of the spectral fitting around
the K$\alpha$ line of He-like Fe, or 6.0--7.2 keV.}
\tablenotemark{d}{Fe abundance derived from the flux ratio of K$\alpha$ line of He-like Fe and the continuum 
at 3.5--6.0 keV.}
\tablenotemark{e}{ICM temperature derived from the spectral fitting within energy range of 0.8--7.3 keV,
considering 10\% systematic error.}
\end{deluxetable}

\begin{deluxetable}{llrrrcrr} 
\tablecaption{Line ratios of the Ly$\alpha$ of H-like Si and S to K$\alpha$ to He-like Fe and  Si/Fe and S/Fe ratios.}
\tablewidth{0pt}
\tablehead{
\colhead{radius} & \colhead{$F_{\rm Si}/F_{\rm Fe}$ \tablenotemark{a}}& \colhead{Si/Fe  \tablenotemark{b}} &\colhead{$\chi^2/d.o.f$\tablenotemark{c}} & \colhead{ $F_{\rm S}/F_{\rm Fe}$ \tablenotemark{a}}& \colhead{S/Fe  \tablenotemark{b}} &\colhead{$\chi^2/d.o.f$\tablenotemark{d}}
\\
\colhead{(kpc/$r_{180<kT>}$)} & & \colhead{(solar ratio)} & & & \colhead{(solar ratio)}
}
\startdata
60--130/0.03-0.06 & 0.33 $\pm$ 0.03 & 0.99 $_{-0.08}^{+0.08}$  & 22/22  & 0.23 $\pm$ 0.02 & 1.02 $_{-0.09}^{+0.09}$  & 80/71 \\
130--220/0.06--0.1 & 0.27 $\pm$ 0.04 & 0.87 $_{-0.14}^{+0.14}$  & 22/22 & 0.23 $\pm$ 0.03 & 1.11 $_{-0.16}^{+0.16}$  & 49/71 \\
220--440/0.1--0.2 & 0.28 $\pm$ 0.03 & 0.88 $_{-0.10}^{+0.10}$  & 16/22  & 0.17 $\pm$ 0.03 & 0.76 $_{-0.14}^{+0.14}$  & 38/46 \\
440--650/0.2--0.3 & 0.37 $\pm$ 0.05 & 1.15 $_{-0.19}^{+0.17}$  & 11/22 & 0.16 $\pm$ 0.05 & 0.69 $_{-0.21}^{+0.21}$  & 54/46 \\
650--1090/0.3--0.5 & 0.26 $\pm$ 0.08 & 0.77 $_{-0.25}^{+0.23}$  & 19/22 & 0.22 $\pm$ 0.06 & 0.97 $_{-0.28}^{+0.26}$  & 45/46 \\
\enddata
\label{tab:sigauss}
 \tablenotemark{a}{
Ratio of flux in units of photons cm$^{-2} \rm{s}^{-1}$ of Ly$\alpha$ line of H-like Si or S to K$\alpha$ line of He-like Fe.}
 \tablenotemark{b}{ The Si/Fe or S/Fe abundance ratios in units of solar ratio derived from the line ratio   using the theoretical expectation  by APEC v2.0.1.}
\tablenotemark{c}{ $\chi^2$ and degrees of freedom of the spectral fitting around
the Ly$\alpha$ line of H-like Si, or 1.8--2.1 keV.}
\tablenotemark{d}{ $\chi^2$ and degrees of freedom of the spectral fitting around
the Ly$\alpha$ line of H-like S, or 2.3--2.8 keV.}
\end{deluxetable}

To derive the strength of the Si line accurately and to
obtain suitable statistics, we fitted the raw (without background subtraction) MOS
and PN spectra  of each annular region outside 60 kpc (0.03$r_{180<\rm kT>}$) out to 1090 kpc (0.5$r_{180<\rm kT>}$) simultaneously with
a sum of the vAPEC code v2.0.1,  two Gaussians for the Ly $\alpha$ line of H-like Si
and the K$\alpha$ line of He-like Si,  a power-law model with $\Gamma=1.4$ for the cosmic
X-ray background (CXB), and a power-law model that is not folded through the 
auxiliary response file (ARF) for the NXB.
Here, we used an energy range of 1.8--2.1 keV, the Si abundance of the
vAPEC model was fixed at 0, and the abundance of the other metals was assumed to
have a same value which was left free.
Above 1.8 keV, a systematic uncertainty caused by a strong instrumental fluorescence line
at $\sim$1.7 keV of the MOS detector does not affect the derived Si abundance.
To derive S and Fe line strengths, we fitted the spectra in the same manner but used
an energy range of 2.3--2.8 keV and 6.0--7.2 keV, respectively.
We added a 6.4 keV Gaussian for an instrumental background line.
We did not used the region within 60 kpc where the temperature structure 
is rather complicated because of  the cool core \citep{Fabian2006}.
Tables \ref{tab:fegauss} and \ref{tab:sigauss} shows the results of the spectral fits.
We obtained reasonable $\chi^2$ values and, 
as shown in Figure \ref{fig:sispec},
the spectra around Si,  S and Fe lines of the MOS and PN were well reproduced by the model.

\begin{figure}[t]
\epsscale{1.0}
\plotone{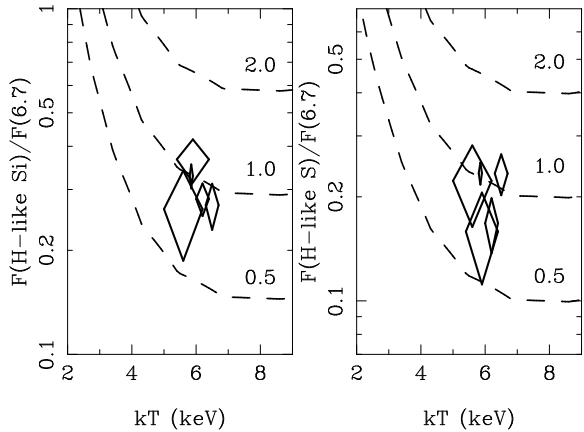}
\caption{ Flux ratios of Ly $\alpha$ line of H-like
Si (left panel) and S (right panel) to K$\alpha$ line of He-like Fe  plotted against plasma temperature.
The temperatures and line ratios were derived within the radial ranges
listed in Tables \ref{tab:fegauss} and  \ref{tab:sigauss}.
Dashed lines  indicate theoretical ratios with a plasma of
constant Si/Fe and S/Fe ratios  according to APEC v2.0.1 model.
The numbers on the plots show the Si/Fe or S/Fe ratios in solar units.
} 
\label{fig:sifelineratio}
\end{figure}
\begin{figure}[t]
\epsscale{1.0}
\plotone{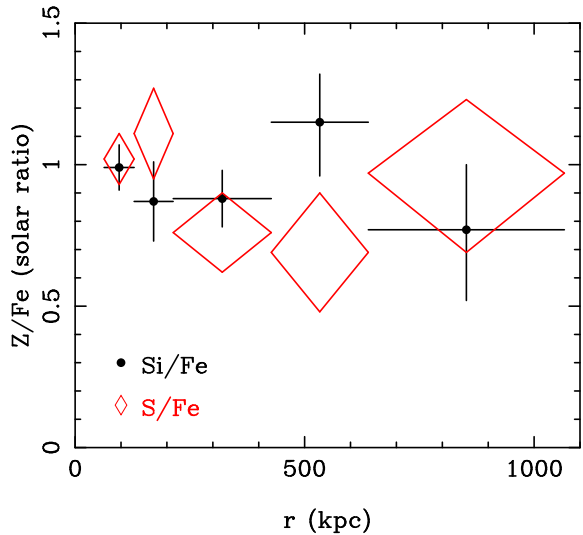}
\caption{Radial profiles of  abundance ratios of Si/Fe (closed circles with error bars) and 
S/Fe (diamonds) in units of solar ratio of the 
Perseus cluster, derived from  flux ratio of Ly$\alpha$ lines of H-like Si or S
to K$\alpha$ line of  He-like Fe.
} 
\label{fig:siferatio}
\end{figure}
\begin{figure}[tbh]
\epsscale{0.9}
\plotone{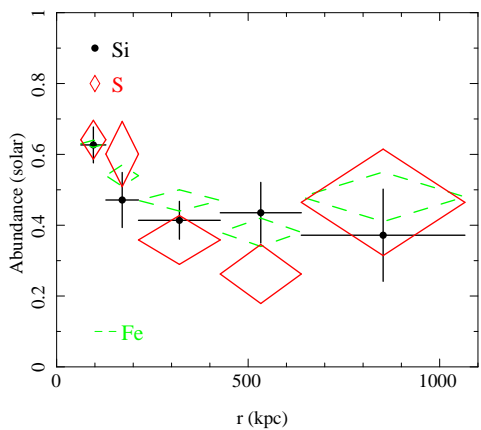}
\caption{ Radial profiles of  abundances of Si (closed circles with error bars),
S (solid diamonds), and Fe (dashed diamonds).
} 
\label{fig:sifeabund}
\end{figure}

Within 0.3$r_{180<\rm kT>}$, we derived the ICM temperature 
from the line ratio of 
Ly$\alpha$ of H-like Fe to K$\alpha$ of He-like Fe using the APEC v2.0.1 
plasma code to reduce systematic uncertainties caused by those in
the response matrix and background.
At 0.3--0.5$r_{180<\rm kT>}$, 
 we used  deep sky observations as a background, 
and fitted the background-subtracted
spectra within an energy range of 0.8--7.3 keV with an APEC model and a
 power-law without ARF for a possible remaining NXB component.
Furthermore, we added  an 10\% systematic error in the derived ICM 
temperature at 0.3--0.5$r_{180<\rm kT>}$ considering the systematic 
uncertainties in the response matrix and background
\citep{Matsushita2011}.
The derived ICM temperatures are shown in Table \ref{tab:fegauss}.

Figure \ref{fig:sifelineratio} shows the 
 ratio of Ly$\alpha$ of H-like Si or S to the K$\alpha$ line of He-like Fe
plotted against the ICM temperature.
The derived line ratios of Si and S to Fe are close to 
theoretical expectations at solar Si/Fe and S/Fe ratios, respectively,
by APEC v2.0.1 (Figure \ref{fig:sifelineratio}).
These two line ratios show a similar temperature dependence: above 5 keV, 
the line ratios are relatively flat at fixed Si/Fe or S/Fe ratios.
The weak temperature dependence can minimize the effect of uncertainties in the temperature
structure of the ICM.
At a given ICM temperature and  abundance ratio,  the  
theoretical expectations obtained by the MEKAL and APEC codes differ by
approximately  10 \%.
The APEC v1.3 code gave almost the same line ratios with APEC v2.0.1.
Therefore, any
 systematic effect due to the plasma codes is expected to be insignificant.

Using the APEC v2.0.1 code, we converted the derived line ratios to the
abundance ratio of Si/Fe and S/Fe 
assuming the single temperature structure.
Table \ref{tab:sigauss} and Figure \ref{fig:siferatio} summarize
the derived Si/Fe and S/Fe ratios.
The derived Si/Fe and S/Fe ratios show
 no radial gradient out to $\sim$ 1100 kpc, or 0.5$r_{\rm 180<kT>}$.
Since the temperature dependence of the  flux ratio of  these lines is relatively flat
above several keV,  the derived abundance ratios should not 
 change as a result of  underestimating the ICM temperature.
In contrast, if the ICM temperature is overestimated 
or if there is an emission component with a  lower temperature, the
 same flux ratio yields a lower ICM abundance.
As a result, the Si/Fe and S/Fe ratios should not be much greater than unity in solar units.
The emission-weighted averages of the Si/Fe and S/Fe ratios outside the cool core,
130--1090 kpc, or 0.06--0.5$r_{180<\rm kT>}$ are 0.91 $\pm$ 0.08 and 0.93 $\pm$ 0.10, respectively,
in units of the solar ratio.

To derive the Fe abundance,  we used the flux ratio of the K$\alpha$ line of He-like Fe
and the continuum at 3.5--6.0 keV (see Matsushita 2011 for details).
Here, we used the APEC v2.0.1 code and the temperatures shown in Table \ref{tab:fegauss}.
The results are shown in Table \ref{tab:fegauss}.
The derived Fe abundances are almost the same as in \citet{Matsushita2011}.
Figure \ref{fig:sifeabund} shows the radial profiles of Si, S, and
Fe abundances.
Beyond 220 kpc (0.1$r_{180\rm<kT>}$) from the center, 
the Si and S abundances have flat radial profiles at $\sim$0.4 solar.


\begin{figure}[t]
\epsscale{0.9}
\plotone{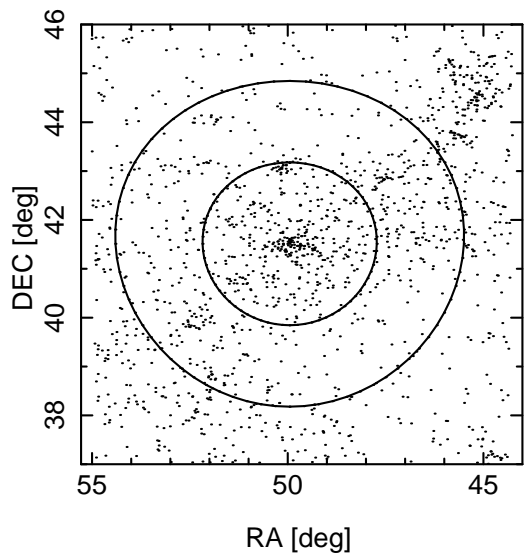}
\caption{ Distribution of galaxies detected by
2MASS in the $K$-band. 
The circles correspond to 1.0 and 2.0$r_{\rm 180<kT>}$.
} 
\label{fig:2massgalaxies}
\end{figure}

\begin{figure}[t]
\epsscale{0.9}
\plotone{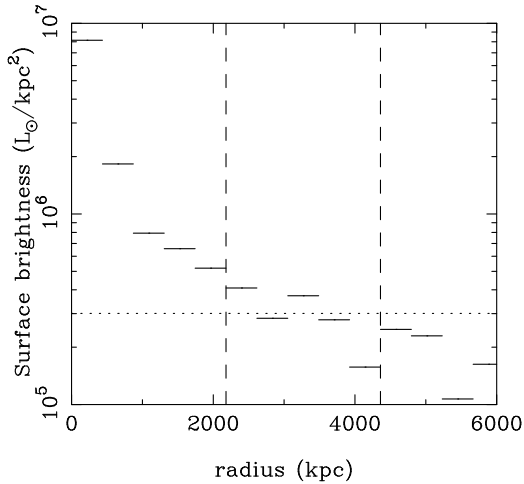}
\caption{
Surface brightness profile of galaxies around the Perseus cluster
observed with 2MASS photometric data in the $K$-band.
Dashed lines
correspond to 1.0$r_{\rm 180<kT>}$ and 2.0$r_{\rm 180<kT>}$.
The dotted line shows the adopted background level.
}
\label{fig:sblk}
\end{figure}

\begin{deluxetable}{rcccc}
\tablecaption{Cumulative values of $K$-band luminosity ($L_K$), gas mass ($M_{\rm gas}$), IMLR, and SMLR derived with XMM.}
\tablewidth{0pt}
\tablehead{
\colhead{radius} & \colhead{$L_K$} & \colhead{$M_{\rm gas}$} & \colhead{IMLR} &  \colhead{SMLR} \\
\colhead{(kpc/$r_{180<\rm kT>}$)} & \colhead{($10^{12} L_\odot$)} & \colhead{($10^{12} M_\odot$)} & \colhead{($10^{-4}M_\odot/L_\odot$)}& \colhead{($10^{-4}M_\odot/L_\odot$)}
}
\startdata
130/0.06 & 2.0 &  3.2  & 12.5 $\pm$ 0.2  & 7.0 $\pm$ 0.1 \\ 
220/0.1  & 2.9 &  6.6  & 16.0 $\pm$ 0.4  & 8.5 $\pm$ 0.2 \\
440/0.2  & 4.3 &  21  &  29 $\pm$ 1   & 15 $\pm$ 1 \\
650/0.3   & 6.0 &  39  &  34 $\pm$ 2 & 20 $\pm$ 1  \\
1090/0.5  & 6.9 & 82  &  64 $\pm$ 5  & 32 $\pm$ 2\\
\enddata
\label{tab:mass}
\end{deluxetable}

\subsection{$K$-band luminosity of galaxies}
\label{subsec:imlr}

\begin{figure}[t]
\epsscale{0.9}
\plotone{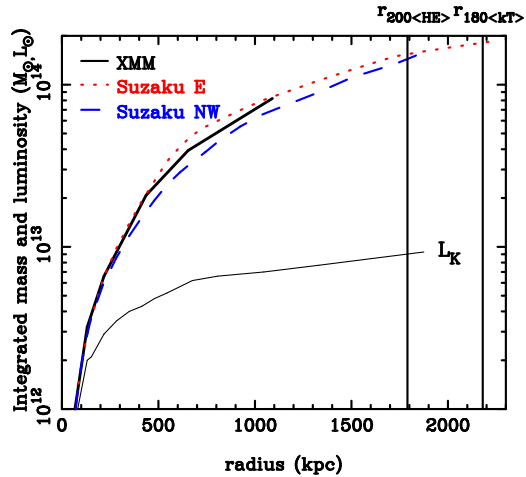}
\caption{ Integrated $K$-band luminosity (thin solid line) and gas-mass profiles of the
Perseus cluster derived with XMM-Newton (bold solid line),
and Suzaku (E:red dotted line, NW: blue dashed line),
integrated over the electron density profiles, 
 assuming spherical symmetry.
The two vertical lines show $r_{180<\rm kT>}$ and $r_{200<\rm HE>}$.
} 
\label{fig:massprofile}
\end{figure}

Because the $K$-band luminosity of a galaxy correlates well with the stellar
mass, we calculated the luminosity profile of the $K$-band.
We collected $K$-band magnitudes of galaxies  in a
$10\times 10~{\rm deg}^2$ box centered on the center of 
the Perseus cluster from the Two
Micron All Sky Survey (2MASS). 
Figure \ref{fig:2massgalaxies} shows  the  galaxies
detected by 2MASS. In the Perseus cluster, the distribution of galaxies 
is elongated in an east-west direction.
 We selected galaxies above the completeness limit of 2MASS, $K_s$ = 13.5 in apparent magnitude\footnote{See http://www.ipac.caltech.edu/2mass/releases/second/doc/explsup.html}.
We corrected the foreground Galactic extinction of $A_\mathrm{K} = 0.06$
 \citep{Schlegel1998} by using the NASA/IPAC Extragalactic Database.
The $K$-band surface-brightness profile of the selected galaxies
 centered on the cD galaxy is shown in Figure \ref{fig:sblk}.
The average surface brightness in the region at
 $1.0r_{180\rm<kT>} < r < 2.0r_{180\rm<kT>}$ ($100' < r < 200'$)  was subtracted as the background.
Next, we deprojected the brightness profile assuming a spherical symmetry and
derived the three-dimensional profile of the $K$-band luminosity.
Figure \ref{fig:massprofile} and Table \ref{tab:mass} show the integrated profile of the $K$-band luminosity.
The total $K$-band luminosity within $r_{180\rm<kT>}$ is  relatively insensitive to the background:
when we used the region 1.0--1.5$r_{180\rm<kT>}$ as the background, the total $K$-band luminosity 
changed by only 10\%.

The completeness limit of galaxies of 2MASS  corresponds to the absolute $K$-band magnitude of -21.0.
This value is much fainter than
the characteristic magnitude, $M_*$ of galaxies in the Perseus cluster 
of -25.09 $\pm$ 0.29
\citep{lin04}.
Integrating the luminosity function with the assumption of the Shechter function,
 the contribution of fainter galaxies below the 2MASS limit is a few \%.

\subsection{Gas-mass, Si-mass, and  Fe-mass profiles}

\begin{figure}[t]
\epsscale{1.0}
\plotone{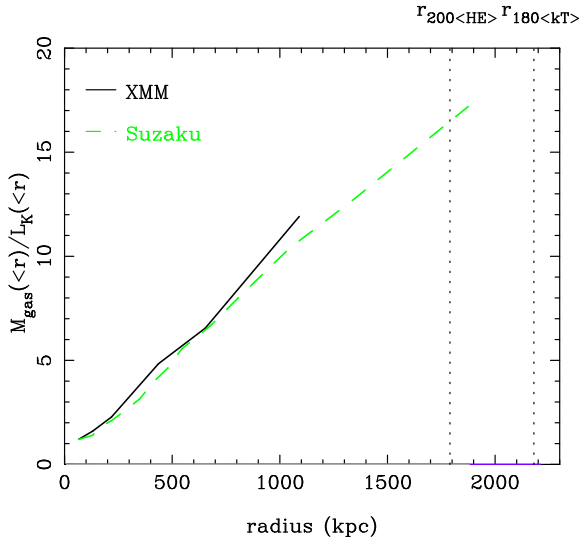}
\caption{ Radial profiles of integrated gas-mass-to-light ratio 
 with XMM (solid line) and Suzaku (dashed line).
Here, we used the weighted average of the E and NW results of Suzaku.
} 
\label{fig:mlr}
\end{figure}

To estimate the integrated gas- and  Fe-mass profiles,
 we accumulated MOS spectra in concentric annular regions
with width of 1$'$--2$'$  out to 50$'$, or 0.5$r_{180<kT>}$.
 We subtracted those of deep sky observations 
accumulated over the same detector regions as background.
We fitted  these spectra  within an energy range of 
1.6--5.0 keV with an APEC model
to avoid uncertainties in the background.
Here, the temperature and Fe abundance were restricted within the error bars
shown in Table \ref{tab:fegauss}.
 We fitted the radial profile of the derived emission measures from the spectral fits per area
with a sum of two $\beta$-models. 
The profile is well represented with the model 
and we calculated the electron density profile.
We integrated the electron density profiles observed with XMM in this way
out to   0.5$r_{180\rm<kT>}$  and
Suzaku toward E and NW directions by
 \citet{Simionescu2011} out to   $\sim r_{200\rm<HE>}$  and 
derived the integrated gas-mass profiles (Figure \ref{fig:massprofile} and
Table \ref{tab:mass}).
Because of  gas sloshing and  cold front at 700 kpc in the E direction,
the integrated gas mass toward this direction 
is higher than that of the NW direction.
The gas-mass profile with XMM falls between the profiles of the two directions
 with
 Suzaku  out to 0.5$r_{180\rm<kT>}$.
With the integrated $K$-band luminosity and gas mass profiles
with XMM and weighted average for the two directions with Suzaku,
 we calculated 
 the  radial profiles of  cumulative  gas-mass-to-light ratio.
In spite of differences in the observed azimuthal directions and
in the analysis methods, 
as shown in Figure \ref{fig:mlr},
the Suzaku and XMM gave similar radial profiles for
the gas-mass-to-light ratio.
The ratio increases with radius out to the virial radius.


\begin{figure}[t]
\epsscale{1.0}
\plotone{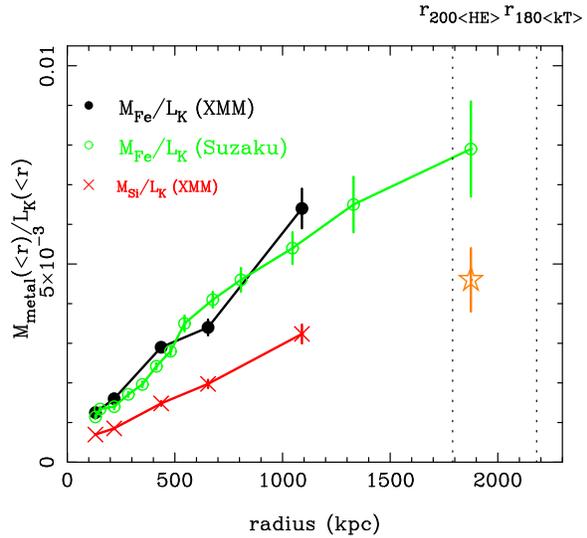}
\caption{ Radial profiles of IMLR with XMM (black closed circles) and Suzaku (green open circles) and SMLR with XMM (red crosses).
Here, we used the weighted average of the E and NW results of Suzaku and assumed 
spherical symmetry.
The orange star indicates the SMLR assuming 
the Si/Fe ratio beyond 0.5$r_{180<\rm kT>}$ is 0.91 $\pm$ 0.08 in solar units.
} 
\label{fig:imlr}
\end{figure}

We used the Si and Fe abundance 
profiles  from XMM-Newton data
and derived integrated mass profiles of Fe and Si in the ICM.
We also integrated the  Fe mass using the 
weighted average of the  Fe abundance profiles for the E and NW directions
observed with Suzaku \citep{Simionescu2011}.
The radial profiles of the cumulative  IMLR and SMLR 
 are shown in Figure \ref{fig:imlr}  and Table \ref{tab:mass}.
The error bars of the mass-to-light ratio include only the errors in abundance.
These profiles continue to increase with radius out to 1800 kpc, $\sim$
$r_{200\rm <HE>}$.

\section{Discussions}

\subsection{The abundance pattern and contribution of SN Ia and SNcc }
The derived Si/Fe and S/Fe ratios in the ICM of  the Perseus cluster
show no radial gradient out to 1.1~Mpc, or 0.5$r_{180\rm<kT>}$.
The emission-weighted averages of the Si/Fe and S/Fe ratios within 0.06--0.5$r_{180\rm<kT>}$ are 0.91 $\pm$ 0.08 and 0.93 $\pm$ 0.10, respectively,
in solar units.
These values are consistent with the emission-weighted average of the Si/Fe ratio 
of 0.99 $\pm$ 0.13 in solar units of the Coma cluster within 0.5$r_{180\rm<kT>}$.
They are also consistent with the values of $\sim$ 0.8 in solar units of
 those in clusters and groups observed with Suzaku 
with ICM temperature less than several keV 
\citep[e.g.][]{matsushita07a,  Sato2007b, sato08,sato09a, sato09b,sato10, komiyama09,   sakuma11}.

 Based on the abundance ratios including Si and Fe, 
the contribution from SN Ia and SNcc were derived 
\citep[e.g.][]{Finoguenov2000, dePlaa2007, Rasmussen2007, Sato2007b,
deGrandi2009, simonescu09, Matsushita2012}.
Since Si and Fe were synthesized  by both SN Ia and SNcc,
the contribution of the two types of SN strongly depends on the adopted nucleosynthesis model
of SN Ia.
With the yields of SNcc with metallicity = 0.02 by \citet{Nomoto2006},
they found that by using  the classical deflagration model, W7 \citep{Iwamoto1999},
and a delayed detonation (DD) model, WDD3 \citep{Iwamoto1999},
for the theoretical SN Ia yield, over a half of the Fe and a few tens of percent of Si 
in the ICM were synthesized in SN Ia.
By adopting another DD model, WDD1 \citep{Iwamoto1999}, 
they found that approximately half of the Si and most of the Fe come from SN Ia.
Suzaku enabled us to measure O and Mg abundances in the ICM outside cool cores of
clusters and groups of galaxies with ICM temperatures smaller than several keV.
\citet{Sato2007b, sato09a, sato10} found that
the mixture of yields of the W7 model and SNcc gave better fits of the observed
 abundance pattern of O, Mg, Si, S, and Fe in the ICM observed with Suzaku
 than for the  WDD1 model and SNcc.
The number ratio of SNcc and SN Ia to
synthesize metals in the ICM  was also  estimated with Suzaku and 
XMM data  \citep[e.g.][]{dePlaa2007, deGrandi2009, simonescu09, Sato2007b, sato09a, sato10}.
Using the Suzaku results including O and Mg,
the number ratio of SNcc to SN Ia was estimated to be $\sim 3.5$ and $\sim 2$,
using the W7 and WDD1 models, respectively.


We also calculated yields mixtures of nucleosynthesis models.
By adopting a Si/Fe ratio of 0.91 $\pm$ 0.08 in solar units in the ICM
of the Perseus cluster, and using the W7 model, we find that
65--74 \% and 23--30 \% of Fe and Si, respectively, were synthesized
by SN Ia.
Using the WDD1 model, 80--91 \% and 56--75 \% of Fe and Si, respectively,
 originated from SN Ia.
Here, we also used the yields of SNcc with metallicity = 0.02 by \citet{Nomoto2006}.
The difference in the Si/Fe ratios in the yields of SNcc assuming a Salpeter 
initial mass function (IMF)
and a top-heavy IMF and different nucleosynthesis models by \citet{Chieffi2004} 
and \citet{Woosley1995} are relatively small, within $\sim$10\%.

From the Si/Fe ratio of the Perseus cluster, we also derived the number
ratio of the SNcc to SN Ia to contribute metals in the ICM.
Using the W7 and WDD1 models, the derived number ratios are
3.2--4.8 and 0.8--2.0, respectively.
The number ratio from the Si/Fe ratio strongly depends on the adopted
nucleosynthesis model, as found by previous studies.

The average S/Si ratio of the ICM in the Perseus cluster is 
1.02 $\pm$ 0.14 in solar units.
The S/Fe ratio of the theoretical yield of SNcc by
\citet{Nomoto2006} with metallicity = 0.02 and the Salpeter IMF
 is 0.83 in solar units.
The yields of the
 W7 model and the WDD models are approximately 1.1--1.2 in solar units.
Therefore,  the observed S/Si ratio of the Perseus cluster is 
 consistent with the  yields of both SN Ia and SNcc.

\subsection{Si-Mass-to-Light Ratios and initial mass function of stars}
\label{subsec:imlr}



The integrated SMLR using the $K$-band
at 0.5$r_{180\rm<kT>}$ of  the Perseus cluster is $\sim$0.003
 $M_\odot/L_{K,\odot}$.
When we assume that the Si/Fe ratio beyond 0.5$r_{180\rm<kT>}$ is the same
as that within the radius, the
 SMLR out to 0.86$r_{180\rm<kT>}$ becomes 0.004--0.005 $M_\odot/L_{K,\odot}$
(Figure \ref{fig:imlr}).
The Si abundance is not expected to increase with radius,
because gas is more extended than stars.
Then, 
assuming the Si abundance beyond 0.5$r_{180}$ is 0.4 solar,
which is the Si abundance at 0.1--0.5$r_{180}$, 
the SMLR becomes 0.005 $M_\odot/L_{K,\odot}$. 
If the clumping of the gas is significant
\citep{Simionescu2011} and the gas fraction at the
virial radius is close to the value of the Wilkinson Microwave Anisotropy Probe (WMAP) 7
\citep{Komatsu2011},
the cumulative SMLR out to 0.86$r_{180\rm<kT>}$ is overestimated
by 20--30\%.
To summarize,  the expected value of the SMLR at 0.86 $r_{180\rm<kT>}$,
is 0.003-0.005 $M_\odot/L_{K,\odot}$.
Adopting the W7 and WDD1 models for the SN Ia nucleosynthesis model,
the SMLR synthesized by SNcc becomes 0.002--0.004  $M_\odot/L_{K,\odot}$ and
0.001--0.002  $M_\odot/L_{K,\odot}$, respectively.


The O, Mg, and Si abundances in the hot interstellar medium (ISM) in  early-type galaxies
reflect stellar metallicity because the ISM is thought to  come from stellar mass loss.
The abundances of these elements in the hot ISM in bright early-type galaxies
 observed with Suzaku
are approximately 0.5--2 solar \citep{Konami2012}.
Extrapolating the observed gradient of the Mg index in optical spectra of
elliptical galaxies, \citet{Kobayashi1999} calculated that the mean stellar metallicity,
[Fe/H] of individual galaxies ranges from -0.8 to +0.3 and  correlates with
stellar luminosity.
They also found that typical [Mg/Fe] is approximately +0.2.
These results indicate that the Si abundance in stars in giant cluster galaxies is
approximately 0.5--2 solar.
Using $K$-band, 
the stellar mass-to-light ratio  in bright early-type galaxies
are approximately unity \citep{Nagino2009}.
As a result, 
 the SMLR trapped in stars should be 0.0005--0.002 $M_\odot/L_{K,\odot}$.
The estimated value of the total SMLR in the Perseus cluster
(i.e., the sum of the SMLR in the ICM and in stars) is 0.004-0.008 $M_\odot/L_{K,\odot}$.

Theoretical models predict that the oxygen-mass-to-light ratio (OMLR) 
and SMLR of a cluster are very  sensitive functions of
the slope of the IMF \citep{Renzini2005}.
Here, the oxygen and silicon mass are
 a sum of that trapped in stars and that in the ICM.
Adopting a Salpeter IMF with a slope of 2.35,  and difference 
in the stellar mass-to-light ratio between the B-band and $K$-band equal 5,
we find that the expected value of
the SMLR from SNcc is $\sim 0.002~ M_\odot/L_{K,\odot}$.
This value is close to the sum of the SMLR in stars and ICM from SNcc, adopting the
nucleosynthesis yield of WDD1 model.
Using the W7 yields, the SMLR from SNcc corresponds to a slope of $\sim$ 2
based on the calculation by \citet{Renzini2005}.
A top-heavy IMF with a slope of 1.35 overproduces metals more than 
that with a factor of 20.
Therefore, the expected value of the SMLR out to the virial radius
does not need the top-heavy IMF.


\subsection{Fe mass and past SN Ia rate}

The solar Si/Fe abundance ratio in the ICM of the Perseus cluster indicates that
most of the Fe was synthesized by SN Ia.
The estimated rate of the current SN Ia rate in present early-type galaxies 
are 0.1--0.3 SN Ia/(100 yr)/(10$^{10} {L_B}_\odot$) \citep{Capp1997, Capp1999, Tura1999, Shar2007, Mann2008}.
We adopted the Fe mass per SN Ia rate by W7 model of 0.75 $M_\odot$
\citep{Iwamoto1999}  and  $L_K/L_B$$\sim 5 {L_K}_\odot/{L_B}_\odot$  for early-type galaxies
\citep{Nagino2009}.
Then,  accumulating the present SN Ia rate
over the Hubble time, 13.7 Gyr, 
the expected IMLR from the SN Ia 
becomes (2--6)$\times 10^{-4} M_\odot/L_{K,\odot}$.
This value is over an order of magnitude smaller than the observed IMLR 
 (6--7) $\times 10^{-3} M_\odot/L_{K,\odot}$ and
 (7--9)$\times 10^{-3} M_\odot/L_{K,\odot}$
within 0.5 $r_{180\rm<kT>}$ and $\sim$0.86 $r_{180\rm<kT>}$, respectively.


The increase of the radial profile of the IMLR  with radius of clusters
 indicates
that Fe in the ICM extends farther than stars, at least out to 0.5 $r_{180\rm <kT>}$.
\citet{leccardi08} and \citet{Matsushita2011} discovered that
the Fe abundance profiles of the ICM within $r_{500}$ are flatter than
expected from
the numerical simulations by \citet{Fabjan2008}, without AGN feedback.
These results indicate that a significant fraction of Fe is 
synthesized in an early phase of cluster evolution,
because if metal enrichment occurs after the formation of clusters,
the metal distribution is expected to follow the stellar distribution.
Considering that most  Fe is synthesized by SN Ia,
the lifetimes of most of SN Ia are much shorter than the Hubble time,
and the SN Ia rate in cluster galaxies was much higher in the past.
If the IMF is close to that in our Galaxy and if most of stars in our Galaxy
and clusters were already formed before a few Gyrs ago, the abundance
pattern (including the ICM and stars in cluster galaxies) 
should naturally be similar to the solar abundance pattern.

\subsection{Comparison of  radial profiles of  IMLR with other systems}
\label{subsec:gmlr}

\begin{figure}[t]
\epsscale{1.0}
     \plotone{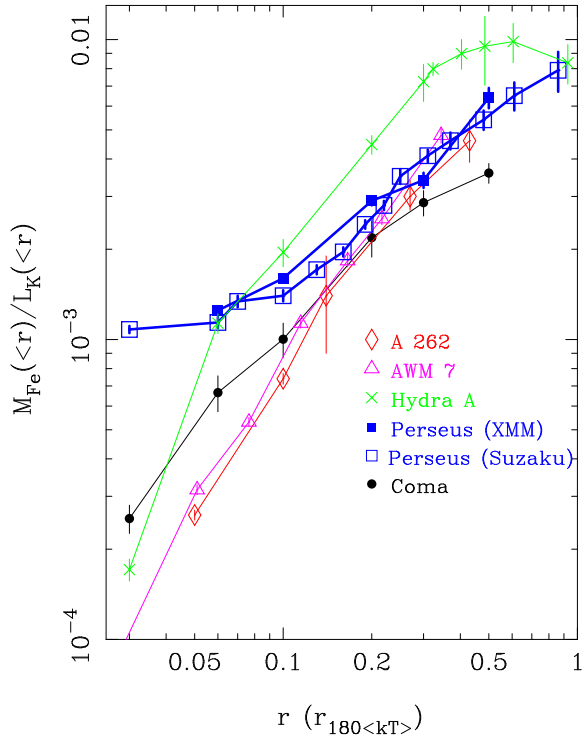}
\caption{
  Profiles of IMLR
plotted against  radius scaled with $r_{180\rm<kT>}$
of the Perseus cluster (closed squares with XMM and open squares with Suzaku),
 AWM 7 (open triangles, \cite{sato08}, Sato et al. in preparation), 
Abell 262 (open diamonds, \cite{sato09b}), 
Hydra A cluster (crosses, \cite{sato12}) and the Coma cluster
(closed circles, \cite{Matsushita2012}).
 }
\label{fig:imlrclusters}
\end{figure}

Figure \ref{fig:imlrclusters} compares the cumulative IMLR profile
 of the Perseus cluster  with that of clusters of galaxies
observed with Suzaku or XMM out to 0.5--1.0$r_{180<\rm kT>}$
including Coma (8 keV; Matsushita et al. 2012),
Abell~262 (2 keV; Sato et al. 2009b),
 AWM 7 (3.6 keV; Sato et al. 2008),
and the Hydra A (3.0 keV; Sato et al. 2012) clusters.
The IMLR profiles of three cool-core clusters at a similar redshift,
Abell 262, AWM 7, and the Perseus cluster agree very well with each other
at 0.2--0.5 $r_{180<\rm kT>}$,
whereas the Coma cluster has a lower IMLR at 0.5 $r_{180<\rm kT>}$.
There is no significant temperature dependence on the derived IMLR profiles.
Since the IMLR profiles increase with radius,
comparison of the total IMLR values  requires 
observations of IMLR profiles of  clusters out to the virial radii.

At a given radius in units of $r_{180<\rm kT>}$,
the Hydra A cluster has a factor-of-two-higher IMLR than the Perseus cluster.
Because of the  higher redshift 
of the Hydra A cluster and lower ICM temperature,
the number of galaxies detected in Hydra A might not be sufficient
 to  deprojected the luminosity profiles in $K$-band.
However, at $\sim r_{180<\rm kT>}$, where systematic uncertainties due
to the limited number of galaxies are relatively small, 
 the cumulative IMLR of the Hydra A 
and the Perseus clusters agree well.
No systematic dependence on the ICM temperature is
 evident for these two clusters
with temperatures of 3 keV and 6 keV.

At the center, within 0.1 $r_{180<\rm kT>}$, the Perseus cluster
and the Hydra A have much
higher IMLR than the other systems, 
which is due to the very bright cool core
of these two  clusters.
Within this radius, the cD galaxies dominate the $K$-band luminosity
and the systematic uncertainties in  the $K$-band luminosity should be small.
The IMLR within  0.1 $r_{180<\rm kT>}$ of the Perseus and
Hydra A clusters are about 1.5--2 $\times 10^{-3} M_\odot/L_{K,\odot}$,
which is a factor 3--10 higher than the expected IMLR from SN Ia
assuming the present SN Ia rate over the Hubble time.
The higher IMLR in these two clusters reflect higher gas-mass-to-light
ratio in 0.1 $r_{180<\rm kT>}$, since the Fe abundance of these two clusters
are not higher than those of the 
other clusters in this region \citep{simonescu09, Matsushita2011}.
\citet{simonescu09} measured the distribution  of O, Si, S, Ar, Ca, Fe
and Ni in the ICM  of the Hydra A cluster within $\sim$0.1 $r_{180<\rm kT>}$.
They found O abundance decreases with radius
and  amount of metals are much higher than the expected values
by stellar winds.
They discussed the initial enrichment by SNcc 
in the early phase in cluster evolution and mixing of these metals.
The O/Fe and Mg/Fe ratios within the cool core of the Perseus cluster
are about unity in solar units \citep{MatsushitaTamura2011} and
agrees with the O/Fe ratio in the cool core of the Hydra A cluster \citep{simonescu09}.
The metals already synthesized in the protocluster phase
can dominate those in the central regions of these two clusters.


\subsection{Gas-mass-to-light ratio}

The integrated gas mass to light ratio of the Perseus cluster
 increases with radius out to 1800 kpc, or $\sim r_{200\rm<HE>}$.
At 1800 kpc, $M_{\rm gas}(<r)/L_K(<r)$ = 16 $M_\odot/L_\odot$.
This value is not so different from 21 $\pm$ 4  $M_\odot/L_\odot$
for the Hydra A cluster ($<kT>$=3 keV) at $r_{200\rm<HE>}$.
The stellar and gas mass fractions within $r_{500}$ depend on the total
system mass \citep{lin03, lin04, Sun09, Gio09, Zhang2010}.
The  gas density profiles in the central regions of groups and poor
clusters were observed  to be shallower than those obtained by  the self-similar model,
and the relative entropy level was correspondingly higher than that
 in rich clusters \citep{Ponman99, ponman03, Sun09}.
Then, the difference in the ratio of gas mass to stellar mass might reflect
differences in distributions of gas and stars, which reflects the history of 
energy injection from galaxies into the ICM.
To study the fractions of stars and gas in clusters of galaxies, and
their dependence on the system mass, 
measurements beyond $r_{500}$ of  other clusters are required.

Based on the Suzaku observations of the NW and E directions of the Perseus cluster,
\citet{Simionescu2011} proposed that the gas-clumping effect is significant 
beyond $r_{500}$, and that the electron density is overestimated.
As a result, $M_{\rm gas}$ within $r_{200\rm<HE>}$ is overestimated
by a factor of 1.3--1.5.
Adopting this value for  the gas mass, the gas-mass-to-light ratio
at  $r_{200\rm<HE>}$ becomes 11-12 $M_\odot/L_\odot$, which is close
to the value at 1000 kpc, or 0.6 $r_{200\rm<HE>}$.
In other words, if there is a significant clumping effect,
 the gas-mass-to-light ratio becomes flat beyond this radius out to $r_{200\rm<HE>}$.


\section{Summary and Conclusion}

We analyzed XMM-Newton data of the Perseus cluster out to 0.5$r_{180<\rm kT>}$
and derived the Si/Fe and S/Fe ratios in the ICM from the flux ratios of the
Ly $\alpha$ lines of H-like Si and S to the K$\alpha$ line of He-like Fe.
The small temperature dependence of the line ratio limits the systematic uncertainty in the derived abundance ratio.
The derived Si/Fe and S/Fe ratios in the ICM show no radial gradient.
The emission-weighted averages of the Si/Fe and S/Fe ratios 
beyond the cool core of the Perseus cluster are 0.91 $\pm$ 0.08 and 0.93 $\pm$ 0.10, respectively,  in solar units.
These abundance ratios indicate that most of Fe in the ICM within 0.5$r_{180<\rm kT>}$ was synthesized by SN Ia.

We collected  $K$-band luminosities in galaxies detected with 2MASS 
and derived the cumulative
radial profile of the $K$-band luminosity of stars in the Perseus cluster.
We calculated the cumulative IMLR and SMLR profiles out to $r_{180<\rm kT>}$.
Furthermore, by using the electron density profiles and Fe abundance profiles
in the two directions observed with Suzaku, 
we calculated the cumulative IMLR profile out to the virial radius.
We also constrained the SMLR value out to this radius.
The SMLR of the Perseus cluster out to the virial radius is significantly smaller
than expected from  the top-heavy IMF, which has a slope of 1.35,
and is  more consistent of a slope of the Salpeter IMF.
With the present SN Ia rate, 
the IMLR within the virial radius is an order-of-magnitude higher than
the SN Ia yield  accumulated over the Hubble time. 
Since the IMLR increases with radius, the most of Fe from SN Ia
were synthesized in an early-phase of cluster formation.


\begin{thebibliography}{}


\bibitem[Anders \& Grevesse(1989)]{angr} Anders, E., \& Grevesse, N.,\ 1989, \gca, 53, 197 

\bibitem[Arnaud et 
al.(1992)]{Arnaud1992} Arnaud, M., Rothenflug, R., Boulade, O., Vigroux, L., \& Vangioni-Flam, E.\ 1992, \aap, 254, 49 


\bibitem[Cappellaro et al.(1997)]{Capp1997} Cappellaro, E., Turatto, M., Tsvetkov, D.~Y., Bartunov, O.~S., Pollas, C., Evans, R., \& Hamuy, M.\ 1997, \aap, 322, 431 
\bibitem[Cappellaro et al.(1999)]{Capp1999} Cappellaro, E., Evans, R., \& Turatto, M.\ 1999, \aap, 351, 459 

\bibitem[Chieffi 
\& Limongi(2004)]{Chieffi2004} Chieffi, A., \& Limongi, M.\ 2004, \apj, 608, 405 

\bibitem[de Grandi \& Molendi(2009)]{deGrandi2009} de Grandi, S., \& Molendi, S.\ 2009, \aap, 508, 565 

\bibitem[de Plaa et al.(2007)]{dePlaa2007} de Plaa, J., Werner, N., Bleeker, J.~A.~M., et al.\ 2007, \aap, 465, 345 


\bibitem[Evrard et al.(1996)]{Evrard1996} Evrard, August E., Metzler, Christopher A. \& Navarro, Julio F. 1996, \apj, 469, 494

\bibitem[Fabian et al.(2006)]{Fabian2006} Fabian, A.~C., Sanders, 
J.~S., Taylor, G.~B., et al.\ 2006, \mnras, 366, 417 
\bibitem[Fabjan et al.(2008)]{Fabjan2008} Fabjan, D., Tornatore, 
L., Borgani, S., Saro, A., \& Dolag, K.\ 2008, \mnras, 386, 1265
\bibitem[Finoguenov et al.(2000)]{Finoguenov2000} Finoguenov, A., David, L. P. \& Ponman, T. J. 2000, ApJ, 544, 188
\bibitem[Finoguenov et al.(2001)]{Finoguenov2001} Finoguenov, A., Arnaud, M. \& David, L. P. 2001, \apj, 555, 191

\bibitem[Fukazawa et al.(1998)]{Fukazawa1998}
Fukazawa, Y., Makishima, K., Tamura, T., Ezawa, H., 
Xu, H., Ikebe, Y., Kikuchi, K. \& Ohashi, T. 1998, \pasj, 50, 187
\bibitem[Fukazawa et al.(2000)]{fukazawa00} Fukazawa, Y., Makishima, K., Tamura, T., Nakazawa, K.,
Ezawa, H., Ikebe, Y., Kikuchi, K. \& Ohashi, T. 2000, \mnras, 313, 21

\bibitem[Fujita et al.(2008)]{fujita08} Fujita, Y., Tawa, N., 
Hayashida, K., et al.\ 2008, \pasj, 60, 343 

\bibitem[Giodini et al.(2009)]{Gio09} Giodini, S., et al.\ 
2009, \apj, 703, 982 










\bibitem[Iwamoto et al.(1999)]{Iwamoto1999} Iwamoto, K., Brachwitz, F., Nomoto, K., Kishimoto, N., Umeda, H., Hix, W.~R., \& Thielemann, F.-K.\ 1999, \apjs, 125, 439














\bibitem[Kobayashi \& Arimoto(1999)]{Kobayashi1999} Kobayashi, C., \& Arimoto, N.\ 1999, \apj, 527, 573 

\bibitem[Komatsu et al.(2011)]{Komatsu2011} Komatsu, E., Smith, 
K.~M., Dunkley, J., et al.\ 2011, \apjs, 192, 18 

\bibitem[Komiyama et al.(2009)]{komiyama09} Komiyama, M., Sato, K., Nagino, R., Ohashi, T. \& Matsushita, K. 2009, \pasj, 61, 337

\bibitem[Konami et al.(2012)]{Konami2012} Konami, S. Matsushita, K., Nagino, R., and Tamagawa, T.  submitted to ApJ




\bibitem[Leccardi \& Molendi(2008)]{leccardi08} Leccardi, A. \& Molendi, S. 2008, \aap, 487, 461

\bibitem[Lin, Mohr, Stanford (2003)]{lin03} Lin, Y.-T., Mohr, J.~J., 
\& Stanford, S.~A.\ 2003, \apj, 591, 749 


\bibitem[Lin \& Mohr(2004)]{lin04} Lin, L. T., \& Mohr, J., J.\ 2004, \apj, 617, 879

\bibitem[Lodders(2003)]{lod03} Lodders, K.\ 2003, \apj, 591, 1220

\bibitem[Makishima et al.(2001)]{Makishima2001} Makishima, K. et al. 2001, \pasj, 53, 401

\bibitem[Mannucci et al.(2008)]{Mann2008} Mannucci, F., Maoz, D., Sharon, K., Botticella, M.~T., Della Valle, M., Gal-Yam, A., \& Panagia, N.\ 2008, \mnras, 383, 1121
\bibitem[Markevitch et al.(1998)]{Markevitch1998} Markevitch, M., Forman, W. R., Sarazin, C. L. \& Vikhlinin, A. 1998, \apj, 503, 77
\bibitem[Matsushita et al.(2003)]{Matsushita2003}
Matsushita, K., Finoguenov, A. \& B{\"o}hringer, H. 2003, \aap, 401, 443
\bibitem[Matsushita et al.(2007a)]{matsushita07a} Matsushita, K. et al. 2007a, \pasj, 59, 327

\bibitem[Matsushita et al.(2007b)]{Matsushita2007b}
Matsushita, K., B{\"o}hringer, H., Takahashi, I. \& 
Ikebe, Y., 2007b, \aap, 462, 953

\bibitem[Matsushita(2011)]{Matsushita2011} Matsushita, K.\ 2011, \aap, 527, A134 

\bibitem[Matsushita et al.(2012)]{Matsushita2012} Matsushita, K.,
Sato, T., Sakuma, E., \& Sato, K., 2012, PASJ  in press, arXiv:1208.609 

\bibitem[Matsushita \& Tamura (2011)]{MatsushitaTamura2011} Matsushita, K. \& Tamura, T. 2011, submitted to A\&A








\bibitem[Nagino 
\& Matsushita(2009)]{Nagino2009} Nagino, R., \& Matsushita, K.\ 2009, \aap, 501, 157 

\bibitem[Navarro et al.(1996)]{navarro1996} Navarro, J.~F., Frenk, 
C.~S., \& White, S.~D.~M.\ 1996, \apj, 462, 563 

\bibitem[Navarro et al.(1997)]{navarro1997} Navarro, J. F., Frenk, C. S., \& White, S. D. M.\ 1997, \apj, 490, 493


\bibitem[Nomoto et al.(2006)]{Nomoto2006}
 Nomoto,~K., Tominaga,~N., Umeda,~H., Kobayashi,~C., 
\& Maeda,~K.\ 2006, Nuclear Physics A, 777, 424 



\bibitem[Ponman, Cannon, Navarro (1999)]{Ponman99}
 Ponman,~T.~J., Cannon,~D.~B., \& Navarro,~J.~F.\ 1999, \nat, 397, 135 

\bibitem[Ponman, Sanderson, Finoguenov(2003)]{ponman03} Ponman, T. J., Sanderson, A. J. R., \& Finoguenov, A.\ 2003, \mnras, 343, 331

\bibitem[Rasmussen \& Ponman(2007)]{Rasmussen2007}
 Rasmussen,~J., \& Ponman,~T.~J.\ 2007, \mnras, 380, 1554 

\bibitem[Rasmussen \& Ponman(2009)]{Rasmussen2009} Rasmussen, J., \& Ponman, T.~J.\ 2009, \mnras, 399, 239 

\bibitem[Renzini et al.(1993)]{Renzini1993} Renzini, A., Ciotti, 
L., D'Ercole, A., \& Pellegrini, S.\ 1993, \apj, 419, 52 
\bibitem[Renzini(2005)]{Renzini2005} 
Renzini, A.\ 2005, The Initial 
Mass Function 50 Years Later, Edited by E. Corbelli and F. Palle, INAF Osservatorio Astrofisico di Arcetri, Firenze, Italy; H. Zinnecker, Astrophysikalisches Potsdam, Germany. Astrophysics and Space Science Library Volume 327. Published by Springer, Dordrecht, 2005, p.221 


\bibitem[Sakuma et al.(2011)]{sakuma11} Sakuma, E., Ota, N., 
Sato, K., Sato, T., \& Matsushita, K.\ 2011, \pasj, 63, 979 


\bibitem[Sato et al.(2007)]{Sato2007b}
Sato, K., Tokoi, K., Matsushita, K., Ishisaki, Y., Yamasaki, N. Y., Ishida, M. \& Ohashi, T. 2007b, \apj, 667, 41
\bibitem[Sato et al.(2008)]{sato08} Sato, K., Matsushita, K., Ishisaki, Y., Yamasaki, N. Y., Ishida, M., Sasaki, S. \& Ohashi, T. 2008, \pasj, 60, 333

\bibitem[Sato et al.(2009a)]{sato09a} Sato, K., Matsushita, K., Ishisaki, Y., Yamasaki, N. Y., Ishida,
M. \& Ohashi, T. 2009a, \pasj, 61, 353

\bibitem[Sato et al.(2009b)]{sato09b} Sato, K., Matsushita, K. \& Gastaldello, F. 2009b, \pasj, 61, 365

\bibitem[Sato et al.(2010)]{sato10} Sato, K., Kawaharada, M., Nakazawa, K., Matsushita, K., Ishisaki, Y., Yamasaki, N. Y. \& Ohashi, T.\ 2010, \pasj, 62, 1445




\bibitem[Sato et al.(2012)]{sato12} Sato, T., Sasaki, T., 
Matsushita, K., et al.\ 2012, \pasj, 64, 95 


\bibitem[Sharon et al.(2007)]{Shar2007} Sharon, K., Gal-Yam, A., Maoz, D., Filippenko, A.~V., \& Guhathakurta, P.\ 2007, \apj, 660, 1165

\bibitem[Schlegel, Finkbeiner, Davis(1998)]{Schlegel1998} Schelegel, D. J., Finkbeiner, D. P., \& Davis, M.\ 1998, \apj, 500, 525

\bibitem[Simionescu et al.(2009)]{simonescu09} Simionescu, A., Werner, N., B{\"o}hringer, H., Kaastra, J. S., Finoguenov, A., Br{\"u}ggen, M. \& Nulsen, P. E. J.\ 2009, \aap, 493, 409


\bibitem[Simionescu et al.(2011)]{Simionescu2011} Simionescu, A., et al.\ 2011, Science, 331, 1576

\bibitem[Simionescu et al.(2012)]{Simionescu2012} Simionescu, A., 
Werner, N., Urban, O., et al.\ 2012, \apj, 757, 182 

\bibitem[Smith et al.(2001)]{smith01} Smith, R. K., Brickhouse, N. S., Liedahl, D. A., \& Raymond, J. C.\ 2001, \apjl, 556, L91 

\bibitem[Sun et al.(2009)]{Sun09} Sun, M., Voit, G.~M., 
Donahue, M., Jones, C., Forman, W., \& Vikhlinin, A.\ 2009, \apj, 693, 1142 

\bibitem[Tamura et al.(2004)]{Tamura2004} Tamura, T., Kaastra, J.~S., den Herder, J.~W.~A., Bleeker, J.~A.~M., \& Peterson, J.~R.\ 2004, \aap, 420, 135 


\bibitem[Tamura et al.(2009)]{tamura2009} Tamura, T., Maeda, Y., 
Mitsuda, K., et al.\ 2009, \apjl, 705, L62 





\bibitem[Tsuru (1992)]{Tsuru1992} Tsuru, T., 1992,  PhD thesis, University of Tokyo


\bibitem[Turatto et al.(1999)]{Tura1999} Turatto, M., Cappellaro, E., \& Petrosian, A.~R.\ 1999, Activity in Galaxies and Related Phenomena, 194, 364









\bibitem[Woosley 
\& Weaver(1995)]{Woosley1995} Woosley, S.~E., \& Weaver, T.~A.\ 1995, \apjs, 101, 181 


\bibitem[Zhang et al.(2010)]{Zhang2010} Zhang, Y.-Y., Okabe, N., 
Finoguenov, A., et al.\ 2010, \apj, 711, 1033 

\end{thebibliography}
\end{document}